\documentclass[useAMS,usenatbib]{mn2e}
\usepackage{tabularx}
\usepackage{xspace}
\usepackage{graphicx}
\newcommand{\ignore}[1]{}
\providecommand{\ao}{}

\renewcommand{\ao}{adaptive optics (AO)\renewcommand{\ao}{AO\xspace}\renewcommand{\Ao}{AO\xspace}\xspace}
\newcommand{\Ao}{Adaptive optics (AO)\renewcommand{\ao}{AO\xspace}\renewcommand{\Ao}{AO\xspace}\xspace}
\newcommand{\wfs}{wavefront sensor (WFS)\renewcommand{\wfs}{WFS\xspace}\renewcommand{\wfss}{WFSs\xspace}\xspace}
\newcommand{\wfss}{wavefront sensors (WFSs)\renewcommand{\wfs}{WFS\xspace}\renewcommand{\wfss}{WFSs\xspace}\xspace}
\newcommand{\shwfs}{Shack-Hartmann \wfs (SHWFS)\renewcommand{\shwfs}{SHWFS\xspace}\xspace}
\newcommand{\dm}{deformable mirror (DM)\renewcommand{\dm}{DM\xspace}\renewcommand{\dms}{DMs\xspace}\renewcommand{\Dms}{DMs\xspace}\renewcommand{\Dm}{DM\xspace}\xspace}
\newcommand{\dms}{deformable mirrors (DMs)\renewcommand{\dm}{DM\xspace}\renewcommand{\dms}{DMs\xspace}\renewcommand{\Dms}{DMs\xspace}\renewcommand{\Dm}{DM\xspace}\xspace}
\newcommand{\Dms}{Deformable mirrors (DMs)\renewcommand{\dm}{DM\xspace}\renewcommand{\dms}{DMs\xspace}\renewcommand{\Dms}{DMs\xspace}\renewcommand{\Dm}{DM\xspace}\xspace}
\newcommand{\Dm}{Deformable mirror (DM)\renewcommand{\dm}{DM\xspace}\renewcommand{\dms}{DMs\xspace}\renewcommand{\Dms}{DMs\xspace}\renewcommand{\Dm}{DM\xspace}\xspace}

\newcommand{\lqg}{linear-quadratic-gaussian (LQG)\renewcommand{\lqg}{LQG\xspace}\xspace}
\newcommand{\shs}{Shack-Hartmann sensor (SHS)\renewcommand{\shs}{SHS\xspace}\renewcommand{\shss}{SHSs\xspace}\xspace}
\newcommand{\shss}{Shack-Hartmann sensors (SHSs)\renewcommand{\shs}{SHS\xspace}\renewcommand{\shss}{SHSs\xspace}\xspace}
\newcommand{\lgs}{laser guide star (LGS)\renewcommand{\lgs}{LGS\xspace}\renewcommand{\Lgs}{LGS\xspace}\renewcommand{\lgss}{LGSs\xspace}\xspace}
\newcommand{\lgss}{laser guide stars (LGSs)\renewcommand{\lgs}{LGS\xspace}\renewcommand{\Lgs}{LGS\xspace}\renewcommand{\lgss}{LGSs\xspace}\xspace}
\newcommand{\Lgs}{Laser guide star (LGS)\renewcommand{\lgs}{LGS\xspace}\renewcommand{\Lgs}{LGS\xspace}\renewcommand{\lgss}{LGSs\xspace}\xspace}

\newcommand{\Ngs}{Natural guide star (NGS)\renewcommand{\ngs}{NGS\xspace}\renewcommand{\Ngs}{NGS\xspace}\renewcommand{\ngss}{NGSs\xspace}\xspace}
\newcommand{\ngs}{natural guide star (NGS)\renewcommand{\ngs}{NGS\xspace}\renewcommand{\Ngs}{NGS\xspace}\renewcommand{\ngss}{NGSs\xspace}\xspace}
\newcommand{\ngss}{natural guide stars (NGSs)\renewcommand{\ngs}{NGS\xspace}\renewcommand{\Ngs}{NGS\xspace}\renewcommand{\ngss}{NGSs\xspace}\xspace}
\newcommand{\mems}{Micro-Electro-Mechanical Systems (MEMS)\renewcommand{\mems}{MEMS\xspace}\xspace}
\newcommand{\snr}{signal to noise ratio (SNR)\renewcommand{\snr}{SNR\xspace}\xspace}
\newcommand{\Moao}{Multi-object \ao (MOAO)\renewcommand{\moao}{MOAO\xspace}\renewcommand{\Moao}{MOAO\xspace}\xspace}
\newcommand{\moao}{multi-object \ao (MOAO)\renewcommand{\moao}{MOAO\xspace}\renewcommand{\Moao}{MOAO\xspace}\xspace}
\newcommand{\mcao}{multi-conjugate adaptive optics (MCAO)\renewcommand{\mcao}{MCAO\xspace}\xspace}
\newcommand{\ltao}{laser tomographic \ao (LTAO)\renewcommand{\ltao}{LTAO\xspace}\xspace}
\newcommand{\cpu}{central processing unit (CPU)\renewcommand{\cpu}{CPU\xspace}\renewcommand{\cpus}{CPUs\xspace}\xspace}
\newcommand{\cpus}{central processing units (CPUs)\renewcommand{\cpu}{CPU\xspace}\renewcommand{\cpus}{CPUs\xspace}\xspace}
\newcommand{\psf}{point spread function (PSF)\renewcommand{\psf}{PSF\xspace}\renewcommand{\psfs}{PSFs\xspace}\renewcommand{\Psf}{PSF\xspace}\xspace}
\newcommand{\psfs}{point spread functions (PSFs)\renewcommand{\psf}{PSF\xspace}\renewcommand{\psfs}{PSFs\xspace}\renewcommand{\Psf}{PSF\xspace}\xspace}
\newcommand{\Psf}{Point spread function (PSF)\renewcommand{\psf}{PSF\xspace}\renewcommand{\psfs}{PSFs\xspace}\renewcommand{\Psf}{PSF\xspace}\xspace}
\newcommand{\fpga}{field programmable gate array (FPGA)\renewcommand{\fpga}{FPGA\xspace}\renewcommand{\fpgas}{FPGAs\xspace}\xspace}
\newcommand{\fpgas}{field programmable gate arrays (FPGAs)\renewcommand{\fpga}{FPGA\xspace}\renewcommand{\fpgas}{FPGAs\xspace}\xspace}
\newcommand{\sor}{successive over-relaxation (SOR)\renewcommand{\sor}{SOR\xspace}\xspace}
\newcommand{\fdpcg}{Fourier domain pre-conditioned gradient (FDPCG)\renewcommand{\fdpcg}{FDPCG\xspace}\xspace}
\newcommand{\map}{maximum a-posteriori (MAP)\renewcommand{\map}{MAP\xspace}\xspace}

\newcommand{\elt}{Extremely Large Telescope (ELT)\renewcommand{\elt}{ELT\xspace}\renewcommand{\elts}{ELTs\xspace}\renewcommand{\eelt}{European ELT (E-ELT)\renewcommand{\eelt}{E-ELT\xspace}\xspace}\xspace}

\newcommand{\elts}{Extremely Large Telescopes (ELTs)\renewcommand{\elt}{ELT\xspace}\renewcommand{\elts}{ELTs\xspace}\renewcommand{\eelt}{European ELT (E-ELT)\renewcommand{\eelt}{E-ELT\xspace}\xspace}\xspace}

\newcommand{\eelt}{European Extremely Large Telescope (E-ELT)\renewcommand{\eelt}{E-ELT\xspace}\renewcommand{\elt}{ELT\xspace}\renewcommand{\elts}{ELTs\xspace}\xspace}

\newcommand{\dugall}{Durham University generalised adaptive optics laser laboratory (DUGALL)\renewcommand{\dugall}{DUGALL\xspace}\xspace}
\newcommand{\fwhm}{full-width at half-maximum (FWHM)\renewcommand{\fwhm}{FWHM\xspace}\xspace}
\newcommand{\wht}{William Herschel Telescope (WHT)\renewcommand{\wht}{WHT\xspace}\xspace}
\newcommand{\emccd}{electron multiplying CCD (EMCCD)\renewcommand{\emccd}{EMCCD\xspace}\renewcommand{\emccds}{EMCCDs\xspace}\xspace}
\newcommand{\emccds}{electron multiplying CCDs (EMCCDs)\renewcommand{\emccd}{EMCCD\xspace}\renewcommand{\emccds}{EMCCDs\xspace}\xspace}
\newcommand{\dasp}{Durham \ao simulation platform (DASP)\renewcommand{\dasp}{DASP\xspace}\renewcommand{\thedasp}{DASP\xspace}\renewcommand{\Thedasp}{DASP\xspace}\renewcommand{\daspcite}{DASP\xspace}\renewcommand{\daspcite}{DASP\xspace}\xspace}
\newcommand{\daspcite}{Durham \ao simulation platform \citep[DASP,]{basden5}\renewcommand{\dasp}{DASP\xspace}\renewcommand{\thedasp}{DASP\xspace}\renewcommand{\Thedasp}{DASP\xspace}\renewcommand{\daspcite}{DASP\xspace}\xspace}
\newcommand{\thedasp}{the Durham \ao simulation platform (DASP)\renewcommand{\dasp}{DASP\xspace}\renewcommand{\thedasp}{DASP\xspace}\renewcommand{\Thedasp}{DASP\xspace}\renewcommand{\daspcite}{DASP\xspace}\xspace}
\newcommand{\Thedasp}{The Durham \ao simulation platform (DASP)\renewcommand{\dasp}{DASP\xspace}\renewcommand{\thedasp}{DASP\xspace}\renewcommand{\Thedasp}{DASP\xspace}\renewcommand{\daspcite}{DASP\xspace}\xspace}
\newcommand{\mpi}{Message Passing Interface (MPI)\renewcommand{\mpi}{MPI\xspace}\xspace}
\newcommand{\smp}{symmetric multi-processing (SMP)\renewcommand{\smp}{SMP\xspace}\xspace}
\newcommand{\svd}{singular value decomposition (SVD)\renewcommand{\svd}{SVD\xspace}\xspace}
\newcommand{\gpu}{graphics processing unit (GPU)\renewcommand{\gpu}{GPU\xspace}\renewcommand{\gpus}{GPUs\xspace}\xspace}
\newcommand{\gpus}{graphics processing units (GPUs)\renewcommand{\gpu}{GPU\xspace}\renewcommand{\gpus}{GPUs\xspace}\xspace}
\newcommand{\fft}{fast Fourier transform (FFT)\renewcommand{\fft}{FFT\xspace}\xspace}
\newcommand{\ifu}{integral field unit (IFU)\renewcommand{\ifu}{IFU\xspace}\xspace}
\newcommand{\darc}{the Durham \ao real-time controller (DARC)\renewcommand{\darc}{DARC\xspace}\renewcommand{\Darc}{DARC\xspace}\xspace}
\newcommand{\Darc}{The Durham \ao real-time controller (DARC)\renewcommand{\darc}{DARC\xspace}\renewcommand{\Darc}{DARC\xspace}\xspace}
\newcommand{\darccite}{the Durham \ao real-time controller \citep[DARC,]{basden9}\renewcommand{\darc}{DARC\xspace}\renewcommand{\Darc}{DARC\xspace}\renewcommand{\darccite}{DARC\xspace}\xspace}
\newcommand{\cots}{commercial off-the-shelf (COTS)\renewcommand{\cots}{COTS\xspace}\xspace}
\newcommand{\rtcp}{real-time control pipeline (RTCP)\renewcommand{\rtcp}{RTCP\xspace}\xspace}
\newcommand{\rms}{root-mean-square (RMS)\renewcommand{\rms}{RMS\xspace}\xspace}
\newcommand{\sFPDP}{serial Front Panel Data Port (sFPDP)\renewcommand{\sFPDP}{sFPDP\xspace}\xspace}
\newcommand{\wpu}{wavefront processing unit (WPU)\renewcommand{\wpu}{WPU\xspace}\xspace}

\newcommand{\rtcs}{real-time control system (RTCS)\renewcommand{\rtcs}{RTCS\xspace}\renewcommand{\rtcss}{RTCSs\xspace}\xspace}
\newcommand{\rtcss}{real-time control systems (RTCSs)\renewcommand{\rtcs}{RTCS\xspace}\renewcommand{\rtcss}{RTCSs\xspace}\xspace}
\newcommand{\eso}{European Southern Observatory (ESO)\renewcommand{\eso}{ESO\xspace}\renewcommand{\theeso}{ESO\xspace}\xspace}
\newcommand{\theeso}{\renewcommand{\theeso}{ESO\xspace}the \eso}
\newcommand{\scao}{single conjugate \ao (SCAO)\renewcommand{\scao}{SCAO\xspace}\renewcommand{\Scao}{SCAO\xspace}\xspace}
\newcommand{\Scao}{Single conjugate \ao (SCAO)\renewcommand{\scao}{SCAO\xspace}\renewcommand{\Scao}{SCAO\xspace}\xspace}
\newcommand{\glao}{ground layer \ao (GLAO)\renewcommand{\glao}{GLAO\xspace}\xspace}
\newcommand{\eagle}{ELT Adaptive optics for GaLaxy Evolution (EAGLE)\renewcommand{\eagle}{EAGLE\xspace}\xspace}
\newcommand{\maory}{multi-conjugate \ao relay for the \eelt (MAORY)\renewcommand{\maory}{MAORY\xspace}\xspace}
\newcommand{\muse}{Multi Unit Spectroscopic Explorer (MUSE)\renewcommand{\muse}{MUSE\xspace}\xspace}
\newcommand{\vlt}{Very Large Telescope (VLT)\renewcommand{\vlt}{VLT\xspace}\xspace}
\newcommand{\ccd}{CCD\xspace}

\newcommand{\eapd}{electron avalanche photodiode\renewcommand{\eapd}{eAPD\xspace}\xspace}
\newcommand{\tmt}{Thirty Metre Telescope (TMT)\renewcommand{\tmt}{TMT\xspace}\xspace}
\newcommand{\lbt}{Large Binocular Telescope (LBT)\renewcommand{\lbt}{LBT\xspace}\xspace}
\newcommand{\xao}{eXtreme \ao (XAO)\renewcommand{\xao}{XAO\xspace}\xspace}

\newcommand{\vla}{Very Large Array (VLA)\renewcommand{\vla}{VLA\xspace}\xspace}
\newcommand{\jwst}{{\em James Webb Space Telescope} \citep[JWST,][]{jwst}\renewcommand{\jwst}{{\em JWST}\xspace}\xspace}
\newcommand{\hst}{{\em Hubble Space Telescope (HST)}\renewcommand{\hst}{{\em HST}\xspace}\xspace}
\newcommand{\ifss}{integral-field spectrographs (IFSs)\renewcommand{\ifss}{IFSs\xspace}\renewcommand{\ifs}{IFS\xspace}\xspace}
\newcommand{\ifs}{integral-field spectrograph (IFS)\renewcommand{\ifss}{IFSs\xspace}\renewcommand{\ifs}{IFS\xspace}\xspace}
\newcommand{\ifus}{integral field units (IFUs)\renewcommand{\ifus}{IFUs\xspace}\xspace}
\newcommand{\mos}{multi-object spectrograph (MOS)\renewcommand{\mos}{MOS\xspace}\xspace}
\newcommand{\goodss}{Great Observatories Origins Deep Survey (GOODS)-S\renewcommand{\goodss}{GOODS-S\xspace}\xspace}
\newcommand{\goods}{Great Observatories Origins Deep Survey (GOODS)\renewcommand{\goods}{GOODS\xspace}\xspace}
\newcommand{\cmos}{complimentary metal-oxide semiconductor (CMOS)\renewcommand{\cmos}{CMOS\xspace}\xspace}
\newcommand{\scmos}{scientific CMOS (sCMOS)\renewcommand{\scmos}{sCMOS\xspace}\xspace}
\newcommand{\aof}{Adaptive Optics Facility (AOF)\renewcommand{\aof}{AOF\xspace}\xspace}
\newcommand{\dsp}{digital signal processor (DSP)\renewcommand{\dsp}{DSP\xspace}\renewcommand{\dsps}{DSPs\xspace}\xspace}
\newcommand{\dsps}{digital signal processors (DSPs)\renewcommand{\dsp}{DSP\xspace}\renewcommand{\dsps}{DSPs\xspace}\xspace}
\newcommand{\capi}{Coherent Accelerator Processor Interface (CAPI)\renewcommand{\capi}{CAPI\xspace}\xspace}
\newcommand{\qe}{quantum efficiency (QE)\renewcommand{\qe}{QE\xspace}\xspace}
\newcommand{\numa}{non-uniform memory access (NUMA)\renewcommand{\numa}{NUMA\xspace}\xspace}
\newcommand{\uav}{unmanned aerial vehicle (UAV)\renewcommand{\uav}{UAV\xspace}\renewcommand{\uavs}{UAVs\xspace}\xspace}
\newcommand{\uavs}{unmanned aerial vehicles (UAVs)\renewcommand{\uav}{UAV\xspace}\renewcommand{\uavs}{UAVs\xspace}\xspace}

\newcommand{\ncpa}{non-common path aberration (NCPA)\renewcommand{\ncpa}{NCPA\xspace}\renewcommand{\ncpas}{NCPAs\xspace}\xspace}
\newcommand{\ncpas}{non-common path aberrations (NCPA)\renewcommand{\ncpa}{NCPA\xspace}\renewcommand{\ncpas}{NCPAs\xspace}\xspace}
\newcommand{\sdk}{software developers kit (SDK)\renewcommand{\sdk}{SDK\xspace}\renewcommand{\sdks}{SDKs\xspace}\xspace}
\newcommand{\sdks}{software developers kits (SDKs)\renewcommand{\sdk}{SDK\xspace}\renewcommand{\sdks}{SDKs\xspace}\xspace}
\newcommand{\dac}{digital to analogue converter (DAC)\renewcommand{\dac}{DAC\xspace}\xspace}
\newcommand{\nda}{non-disclosure agreement (NDA)\renewcommand{\nda}{NDA\xspace}\xspace}
\newcommand{\polc}{pseudo-open-loop control (POLC)\renewcommand{\polc}{POLC\xspace}\xspace}
\newcommand{\udp}{User Datagram Protocol (UDP)\renewcommand{\udp}{UDP\xspace}\xspace}
\newcommand{\ags}{artificial guide star (AGS)\renewcommand{\ags}{AGS\xspace}\xspace}
\newcommand{\est}{European Solar Telescope (EST)\renewcommand{\est}{EST\xspace}\xspace}
\newcommand{\lot}{Large Optical Telescope (LOT)\renewcommand{\lot}{LOT\xspace}\xspace}
\newcommand{\gtc}{Gran Telescopio Canarias (GTC)\renewcommand{\gtc}{GTC\xspace}\xspace}

\newcommand{\cog}{centre of gravity (COG)\renewcommand{\cog}{COG\xspace}\xspace}
\usepackage{amssymb}

\title[On-sky matched filtering]{On-sky demonstration of
  matched filters for wavefront measurements using ELT-scale elongated laser
  guide stars}

\ignore{
  Iguidoli@eso.org,  Eric Gendron <eric.gendron@obspm.fr>,
  fabrice.vidal@obspm.fr,    Gerard Rousset <gerard.rousset@obspm.fr>,
  dbonacci@eso.org,    mauro.centrone@oa-roma.inaf.it, mreyes@iac.es,
  JENKINS D.R. <d.r.jenkins@durham.ac.uk>,    Richard Myers
  <r.m.myers@durham.ac.uk>, t.j.morris@durham.ac.uk,    dbgh26
  <james.osborn@durham.ac.uk>, d70j6c <a.p.reeves@durham.ac.uk>,
  xfzp51 <matthew.townson@durham.ac.uk>,
      lisa.bardou@obspm.fr,
  damien.gratadour@obspm.fr,
  fanny.chemla@obspm.fr,jean-tristan.buey@obspm.fr,jean-luc.gach@lam.fr, emarchet@eso.org
Gianluca Lombardi, gianluca.lombardi@gtc.iac.es

  }

\author[A.\ G.\ Basden et al.]{A.\ G.\ Basden$^{1}$\thanks{E-mail:
    a.g.basden@durham.ac.uk (AGB)},
  L.\ Bardou$^2$,
  D.\ Bonaccini Calia$^3$,
  T.\ Buey$^2$,
  M.\ Centrone$^4$,
  \newauthor
  F.\ Chemla$^5$,
  J.\ L.\ Gach$^6$,
  E.\ Gendron$^2$,
  D.\ Gratadour$^2$,
  \newauthor
  I.\ Guidolin$^3$,
  D.\ R.\ Jenkins$^1$,
E.\ Marchetti$^3$,
  T.\ J.\ Morris$^1$,
  R.\ M.\ Myers$^1$,
  \newauthor
  J.\ Osborn$^1$,
  A.\ P.\ Reeves$^1$,
  M.\ Reyes$^7$,
  G.\ Rousset$^2$,
  \newauthor
  G.\ Lombardi$^{8}$,
  M.\ J.\ Townson$^1$,
  F.\ Vidal$^2$, \\
  $^{1}$Department of Physics, Durham University, South Road, Durham, DH1 3LE, UK\\
  $^2$ LESIA, Observatoire de Paris - Paris Sciences et Lettres, CNRS, Universit\'e P. et M. Curie - Sorbonne Universit\'es, \\Universit\'e Paris Diderot - Sorbonne Paris Cit\'e, 5, Place Jules Janssen, 92190 Meudon, France\\
  $^3$ European Southern Observatory, Garching, Germany\\
  $^4$ INAF - Instituto Nazionale de Astrofisica, Observatorio
  Astronomico di Roma, Via Frascati 33,
  00078, Monte Porzio Catone (RM), Italy\\
$^5$ GEPI, Observatoire de Paris - Paris Sciences et Lettres, CNRS,
  Universit\'e Paris Diderot - Sorbonne Paris Cit\'e, 5, \\Place Jules
  Janssen, 92190 Meudon, France\\
  $^6$ Laboratoire d’Astrophysique de Marseille, 38 rue
  F. Joliot-Curie, 13388 Marseille Cedex 13, France\\
  $^7$ Instituto de Astrofisica de Canarias, Via Lactea, La Laguna,  Tenerife, Spain\\
  $^8$ Gran Telescopio Canarias, Cuesta de San José, s/n - E-38712, Breña Baja, La Palma, Spain}

\begin{document}
\maketitle

\begin{abstract}
The performance of adaptive optics systems is partially dependant on
the algorithms used within the real-time control system to compute
wavefront slope measurements.  We demonstrate use of a matched filter
algorithm for the processing of elongated laser guide star (LGS)
Shack-Hartmann images, using the CANARY adaptive optics instrument on
the 4.2~m William Herschel Telescope and the European Southern
Observatory Wendelstein LGS Unit placed 40m away.  This algorithm has
been selected for use with the forthcoming Thirty Meter Telescope, but
until now had not been demonstrated on-sky.  From the results of a first observing run, we show that the use of
matched filtering improves our adaptive optics system performance,
with increases in on-sky H-band Strehl measured up to about a factor
of 1.1 with respect to a conventional centre of gravity approach.  We
describe the algorithm used, and the methods that we implemented to
enable on-sky demonstration.

\end{abstract}


\section{Introduction}
\Ao systems have reached a new level of maturity in recent years, with
many new strategies and algorithms being demonstrated and realised in
facility class systems, e.g.\ \xao \citep{sphere,2015PASP..127..890Jshort,2012SPIE.8446E..1UMshort} and \mcao
\citep{2012SPIE.8447E..0IRshort}.  Future systems for the forthcoming \elts are currently
under development, and it is important that optimum performance is
obtained for these expensive facilities.  A key area of investigation
is optimisation of algorithms that are used in the \ao real-time
pipeline, to extract maximum information from the photons received by
the \wfss, and using {\it apriori} knowledge of the atmosphere
to compute the best match of \dm surface to the phase perturbations
introduced by atmospheric turbulence.

In order to increase sky coverage of \ao corrected observations (which
are limited by the availability of bright natural guide stars), \lgss
are used \citep{laserguidestar}.  Artificial stars can be created in
the mesosphere, at about 90~km above the surface of the Earth, using a
laser tuned to the resonance of sodium atoms.  However, due to the thickness of the sodium layer, this leads to
an elongation of the spots in a Shack-Hartmann \wfs when sub-apertures
view the \lgs off-axis, due to perspective \citep{matchedfilter1}.
This elongation is approximately proportional to the distance between
the sub-aperture (projected onto the telescope pupil) and the laser
launch location (projected into the plane of the primary mirror).  For
the \eelt, this distance can reach about 40~m
\citep{doi:10.1117/12.2058467} (for a small number of most affected
sub-aperture), and so spots extend for up to 20--30~arcseconds along
the elongation direction, depending on the precise distribution of
mesospheric sodium atoms.  Spots that are this large introduce several
challenges for wavefront sensing: flux is spread over a larger number
of pixels and so \snr is reduced, and sensitivity to spot motion along
and perpendicular to the elongation direction differs.

Here, we consider the use of a noise-optimal matched filter algorithm
\citep{matchedfilter1,2008OptL...33.1159G,2009ApOpt..48.1198C} for
calculation of local wavefront gradients in a Shack-Hartmann \wfs.  We
report on successful on-sky operation of this algorithm, taking
advantage of a dedicated experiment set to evaluate the performance of
AO using \elt-scale elongated \lgss \citep{canaryD}.  This experiment
is based on the use of the CANARY \ao technology demonstrator
instrument \citep{canaryshort,canaryresultsshort} at the 4.2~m William
Herschel Telescope coupled to the \eso Wendelstein LGS Unit (WLGSU)
\citep{wlgsu} with the laser being launched 40~m away from the WHT.
Performance of the \ao system when using the matched filter
algorithm is compared with that obtained using a well optimised \cog
approach (see \S\ref{sect:cogOptimzed} for further details about the
optimisation).  The matched filter algorithm has been selected for use
with the forthcoming Thirty Meter Telescope \citep{tmt}, and, as we
show, also has relevance for other telescopes.  Previous authors have
also compared the performance of different algorithms used for
wavefront slope determination for Shack-Hartmann \wfss
\citep{2006MNRAS.371..323T,basden18}.  Here we provide results of the first on-sky
verification of one of these.

The CANARY \ao system has seen several phases of operation using many
different \wfss and \dms \citep{basden22short}.  Here, we
operate in phase D mode \citep{canaryD}, with a single on-axis sodium
\lgs, an on-axis \ngs truth \wfs, and optionally, three
off-axis \ngss (which we do not consider further here).  Phase D of
CANARY is designed to emulate a sub-pupil of the \eelt, with the \lgss
being launched far off axis so that significant spot elongation
effects can be studied.  The \wfss all have $7\times7$ sub-apertures, of which 36 are valid (not
significantly vignetted).  The \lgs \wfs has $30\times30$
pixels per sub-aperture, and is based on an \emccd detector with
sub-electron readout noise \citep{ocams}. It is equipped with a dedicated fast steering mirror to stabilize the \lgs spots in the subapertures.  Wavefront correction is performed using
an $8\times8$ actuator \dm with 52 active actuators and a dedicated tip/tilt mirror.

\begin{figure}
  \includegraphics[width=\linewidth]{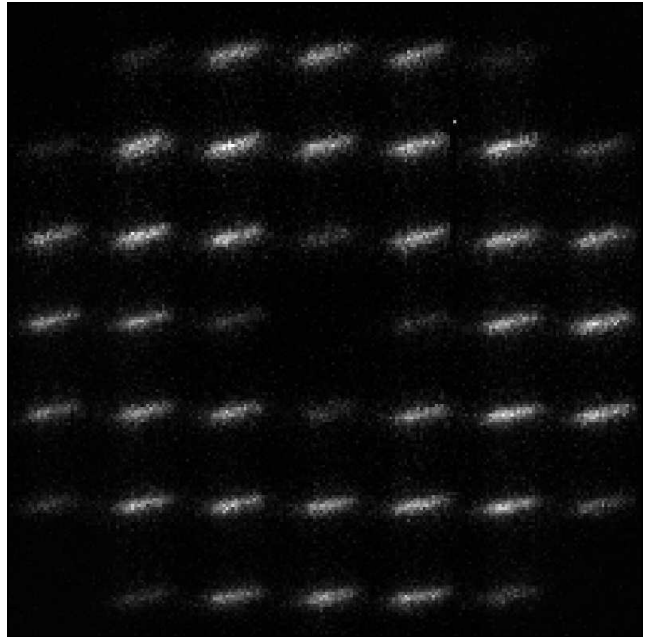}
\caption{A figure showing a single \wfs frame of elongated
  laser guide star spots recorded by the CANARY AO system on 20$^{th}$
  September 2016.  This image has $30\times30$ pixels per
  sub-aperture, and a pixel scale of 0.65~arcseconds per pixel.}
\label{fig:lgsspots}
\end{figure}

We concentrate our efforts on matched filtering of highly elongated
sodium \lgs spots, as shown in Fig.~\ref{fig:lgsspots}, though we have
also tested with \ngs spot patterns.  However, on-sky observing time
was limited and so we do not present results for \ngss: we did not
take comprehensive \ngs measurements once the \ao loop was successfully closed.

The matched filter algorithm has been developed to help mitigate some
of the problems introduced by guide star elongation, including
reduction in flux per pixel and sensitivity loss due to structure.
Previous studies, e.g.\ \citet{matchedfilter1} have found that spot
position estimation error is significantly impacted by this
elongation, and that using a matched filter can reduce this effect,
particularly in the presence of low photon signal and high detector
readout noise.

Unfortunately, the matched filter algorithm is non-linear,
i.e. estimated position and hence \rms error depend non-linearly on
distance of the spot from the reference position of the matched
filter.  However, a linearity extension has been proposed
\citep{2008OptL...33.1159G} which we use to extend the linearity of
this algorithm up to about 1 pixel of displacement.  Additionally,
because we operate in closed loop (as with the \tmt), the \ao system
helps to maintain the spots in a central position, though we note that
spot motion can still be large for wide-field \lgs systems and \glao
systems.  The pixel scale of the \wfs that we use is also fairly
large, at 0.65~arcseconds per pixel, principally so that a reasonable
sub-aperture size can have sufficient field of view to image the
entire \lgs spot, and therefore, under reasonable seeing conditions
(e.g.\ 0.7~arcseconds), \lgs spot motion is small, within the linear range
of the matched filter algorithm.

In the case of the forthcoming \eelt, laser guide stars are launched
from the edge of the telescope pupil, and so the degree of elongation
varies across a Shack-Hartmann \wfs from close to zero
(for sub-apertures near the launch location) to about 40~m (for
sub-apertures furthest from the launch location), which corresponds to
tens of arcseconds of angular diameter for typical sodium layer
profiles and thickness.  In this case, a \cog algorithm can be expected to perform
well for sub-apertures with little elongation, and poorly for well
elongated sub-spertures.  The matched filter algorithm is able to
improve measurement accuracy for all sub-apertures, with the largest
gain being for those that are most elongated.

In \S2 we describe the technique that we use to build and apply the
matched filters.  In \S3 we present on-sky results, and we conclude in \S4.

\section{Matched filters for wavefront gradient estimation}
\label{sect:mf}
We follow the derivation of the matched filter as given by
\citet{matchedfilter1}, with an extension for dynamic range
\citep{2008OptL...33.1159G}.  This is outlined as follows.

The sub-aperture pixel intensity is modelled as:
\begin{equation}
I_{m}\left( \vec{x} \right) = I_0\left( \vec{x} \right) + G\cdot \vec{x} + \eta
\label{eq:model}
\end{equation}
where $I_m$ is the intensity of a pixel at location $\vec{x}$, $I_0$ is
the high light level on-axis intensity (assumed to be noiseless), $G$ is the derivative
of $I_0$ in the $x$ and $y$ directions (the pixel gain, or Jacobian)
having components $G_x$ and $G_y$,
and $\eta$ is the measurement noise vector due to the photon shot
noise and detector readout noise.

The minimisation of the difference between measured pixel intensity
and model pixel intensity (Eq.~\ref{eq:model} is required to compute the estimate of the
spot location, and the derivation is given by
\citet{2008OptL...33.1159G}.  In summary (with small modifications
made by us to allow for variations in flux measurements), estimated
spot position is given by
\begin{equation}
  \vec{s_x} = \frac{R\cdot I}{\sum I}
\end{equation}
where $\vec{s_x}$ is the desired slope measurement, $I$ is the
sub-aperture image (flattened into a one-dimensional array) and $\sum
I$ is a measure of the total flux within $I$. $R$ (the matched
filter), which has both $x$ and $y$ components is given by
\begin{equation}
  R=fM(H^TC^{-1}H)^{-1}H^TC^{-1}
\end{equation}
with $C$ being the noise covariance matrix (a diagonal matrix of side
length equal to the number of pixels within a sub-aperture), with
entries equal to $I_0 + \sigma^2$ for photon and detector noise where
$\sigma$ is the detector readout noise, $f$ is the sum of flux in
$I_0$,
\begin{equation}
 M=  \left[ \begin{array}{ccccccc}
  1& 0& 0& 1& -1& 0& 0 \\
  0& 1& 0& 0& 0& 1& -1 \end{array}\right]
\end{equation}
and
\begin{equation}
H= \left[ \begin{array}{ccccccc} G_x & G_y & I_0& I_1& I_2& I_3& I_4 \end{array}\right]
\end{equation}
Here, $I_{1,2,3,4}$ are the high light level intensities of the spot,
shifted in the +x, -x, +y and -y directions respectively by one pixel.

\subsection{Calculation of the matched filter}
Ideally, calculation of the matched filter would be an ongoing
process, continually computed by dithering of the \lgs spots on the
\wfs, using a tip-tilt mirror \citep{matchedfilter1}, to
allow ongoing calculation of the image gradients, $G_x$ and $G_y$.
Dithering would have little effect on the \ao correction since global
\lgs motion is ignored, being removed from the slope measurements
before being passed to the reconstructor.  Continual calculation means
that the matched filter remains relevant over long periods while
atmospheric conditions, and sodium layer profile change.  The
introduction of a small, known, time-varying astigmatism will enable
continual calculation of the relative pixel scales of both the \lgs
and \ngs \wfss (key to maintaining \ao performance in varying seeing
conditions).  If the introduced astigmatism is kept small (and
detected using a phase locked loop), effect on image \psf will be
minimal.

However, since CANARY is an \ao demonstrator, and does not need to
produce astronomical science, and therefore does not require \ao
stability for long integration periods, we have taken the approach of
building the matched filter off-line to simplify on-sky operation,
which is used for all results presented here.  We hope to develop an
online calculation capability for future sky tests and operation.

Our sequence of operations to build the matched filter is as follows.
\begin{enumerate}
\item The \lgs fast steering mirror is set to dither, with four
  separate positions per cycle.
\item \wfs slope measurements (using a \cog) are
    recorded and used to determine the relative phase difference
    between the mirror and \wfs.  Since this is performed
    on averaged images, signal levels can be high, allowing the \cog
    estimation to be accurate.
\item The phase of the dithering is adjusted so that the shift in
  \wfs spots is aligned with the detector pixels and sub-apertures.
\item A number of \wfs images are recorded, and those
  corresponding to each of the four dithered mirror locations are averaged
  together.  Typically 5--10 seconds of images are recorded.
\item The \cog of these four averaged images is computed, to give the
  dither radius (in pixels).
\item The image gradient, $G_x$ and $G_y$ are computed from the
  difference of opposing pairs of images divided by the diameter of
  the dither (in pixels).
\item The dithering is removed, and typically 5--10 seconds of images
  are averaged together to obtain $I_0$.  $I_1$, $I_2$, $I_3$ and
  $I_4$ are also computed.
\item The matched filter, $R$ is then computed.
\end{enumerate}

When computing the matched filter, we find that tuning the \wfs
noise level estimate, $\sigma$, can be used to improve
performance, since this parameter acts to regularise part of the
solution for $R$.  We use a single value for $\sigma$ rather than a
separate one for each pixel, partly for simplicity, and also because
for a \ccd detector, this is typically constant.  However we note that
for a \scmos detector (or indeed a multi-port \ccd detector) there may
be performance improvements if a different value for each pixel can be
used, i.e.\ less weight can be applied to noisier pixels.

Once the matched filter has been computed, it is uploaded to the
real-time control system and matched filtering switched on.  It is
then necessary for us to measure reference slopes, which we do by
averaging the computed wavefront slope measurements over a sufficient
number of frames for wavefront turbulence to average out.  Typically,
we record for about 60 seconds.  We note however that this method of
removing static wavefront offsets is impractical for larger telescope
apertures, since the time to average turbulence greatly increases \citep{Gordon11}.

\subsection{Application of the matched filter}
The CANARY instrument uses a \cpu-based real-time control system,
\darc \citep{basden9} which provides the necessary flexibility to
implement, develop and test new algorithms.  Within the real-time
control system, the matched filter is applied as follows:
\begin{enumerate}
\item Sub-aperture flux is computed in units of analogue to digital
  units (not photons), i.e.\ the calibrated pixels are summed.
\item The dot product of $R_x$ with the sub-aperture pixels (flattened
  into a vector) is computed, and the result divided by the
  sub-aperture flux to give the $x$ slope estimate.
\item Similarly for the $y$ slope estimate using $R_y$.
\end{enumerate}

\darc has the ability to actively track Shack-Hartmann spots,
i.e.\ the sub-aperture positions can follow the spots, such that the
spots remain close to the centre of the sub-aperture, based on the
location of the spot in the previous image frame.  This therefore
aids the linearity of the matched filter algorithm, since the matched
filter will be applied to spots that are close to the sub-aperture
centre.  The alternative to this approach, which we do not explore, is
a two-step algorithm, which initially estimates spot position with a
\cog algorithm, and then applies the matched filter to this.  However,
this introduces additional computational complexity.

Within \darc, we use a different matched filter for each sub-aperture,
and internally, these are stored in an two-dimensional array where the
location of the matched filter corresponds with the default
sub-aperture position.  Therefore we can apply matched filtering to
sub-apertures of different sizes, and to different spot elongations
across the pupil.

\subsection{Development and testing of matched filters}
\begin{figure}
  \includegraphics[width=\linewidth]{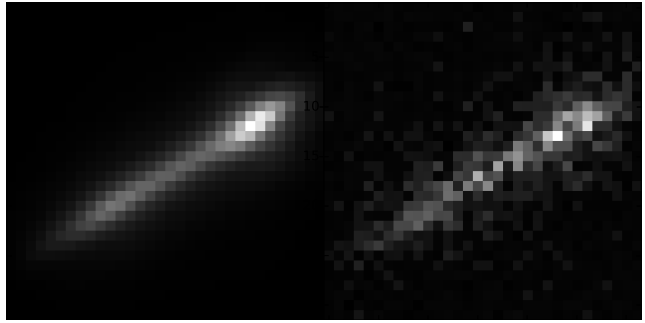}
  \caption{An example atmospherically propagated sodium LGS
    Shack-Hartmann spot, as used in simulations (6.4~arcsecond field
    of view). Left: Noiseless image.  Right: With added photon show
    noise (for a total flux of 1000 photons) and readout noise (0.1
    electrons per pixel \rms).}
  \label{fig:singleshs}
\end{figure}

To develop the real-time implementation of the matched filter
algorithm, and to test the technique used to build the matched filter
we employed two simulation concepts.  First, we use \dasp
\citep{basden5} to model a single sub-aperture with an extended laser
guide star spot viewed through atmospheric turbulence.  We use this
simulation tool to build the matched filter as outlined in
\S\ref{sect:mf}.  We then generate a sequence of atmospherically
propagated Shack-Hartmann spots, as shown in Fig.~\ref{fig:singleshs}
and compare the \cog and matched filter spot position estimators with
the true known location (measured on a noiseless image using a \cog).
The \rms deviation of each estimator from the known position is then
computed.  We are able to explore matched filter performance covering
a parameter space including elongation, signal level, readout noise
level and number of frames used when building the matched filter.
Although we do not seek to present significant simulated results here,
because matched filter performance has been studied elsewhere, for
example \citet{2009ApOpt..48.1198C}, we present a summary of our
modelling.  This modelling is based on a \lgs launched 40~m from the
telescope aperture (approximating CANARY and furthest sub-apertures of
the \eelt), and a 60~cm sub-aperture.  We note that for the \tmt,
which uses centre-launched \lgss, maximum sub-aperture distance from
the launch axis will be about 15~m.  Default photon flux in our
simulations is 1000 photons per sub-aperture per frame
\citep[pessimistic compared with on-sky measurements by a factor of
  about 2--5,][]{lgsflux} with sub-electron readout noise (i.e.\ an
electron multiplying \ccd).  Seeing of 1~arcsecond is assumed, and by
default we use 1000 images to build the matched filters.

After verification of the matched filter algorithm, we proceeded to
develop the real-time implementation and support tools.  To verify the
operation of these, we use the real-time simulation concept
\citep{basden13} coupling a Monte-Carlo simulation tool (\dasp) to the
real-time control system used by CANARY.  This concept is simple:
instead of using real physical cameras and \dms, they are instead
simulated, with the simulation \wfs camera data being fed
into the real-time control system.  The \dm command output is
captured and used to shape simulated \dms.  We are then able to use
the user interface tools developed for on-sky operation with this
simulated system, and therefore provide a thorough testing,
significantly reducing necessary on-sky commissioning time.  The
real-time control system itself does not need to know whether it is
operating with real or simulated cameras and \dms.

We note that since the CANARY real-time control system is \cpu-based,
we do not need to operate the real-time simulation facility on the
same computational hardware, i.e.\ we can, for example, operate the
simulation and real-time control system on a developer's PC, to aid
development when not present at the telescope, and to allow multiple
instances of the system to be run simultaneously.

By using this real-time simulation, we are then able to test building
of the matched filter (including dithering), and application of the
matched filter, including engaging the \ao loop.

\section{Matched filtering results}

We performed tests of matched filtering using the CANARY \ao on-sky
demonstrator instrument over a period of three nights in September
2016.  During this operation, the WLGSU launch telescope was set about
40~m from the \wht to generate the sodium \lgs in the mesosphere.
Therefore, spots were highly
elongated, as shown in Fig.~\ref{fig:lgsspots}.

\subsection{Simulation measurements}
\begin{figure}
  \includegraphics[width=\linewidth]{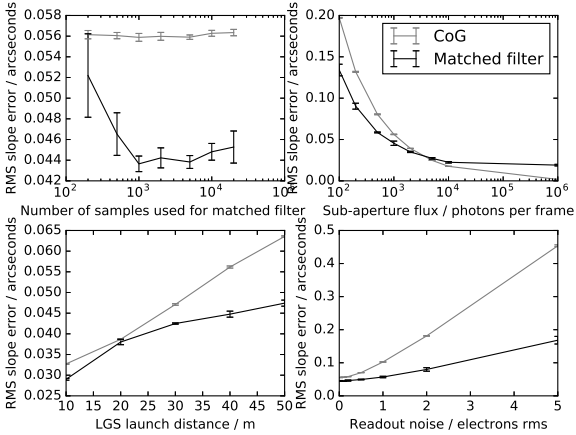}
  \caption{A figure comparing matched filtering with \cog.  The
    sub-plots give performance as \rms estimation slope error (in
    arcseconds) with respect to a noiseless CoG measurement (lower
    error is better performance).  These plots show performance (a) as
    a function of number of images used to build the matched filter,
    (b) as a function of guide star flux, (c) as a function of
    distance of the LGS from the telescope axis, and (d) as a function
    of detector readout noise.}
  \label{fig:simresults}
\end{figure}

Exploration of the matched filter parameter space was used to confirm
the range of expected parameters over which matched filtering would
out perform a \cog algorithm.  Fig.~\ref{fig:simresults}
shows some of these results that we obtained with simulation, and it
can be seen (in agreement with previous studies) that matched
filtering can yield improved performance when compared with \cog.  We
note that at high light levels, the \cog performs better
since we are using a noiseless \cog measurement as the ``truth''.

During \ao operation, if the matched filter is to be updated,
reference slopes will also require updating.  Within simulation we
developed a technique to compute updated reference slopes:
\begin{enumerate}
\item Compute \cog and matched filter measurements of the reference
  image, $I_0$, giving $c_x$, $c_y$, $m_x$, and $m_y$ respectively,
  defined in pixels offset for the $x$ and $y$ directions.
\item Add the matched filter estimates, $m_x$ and $m_y$, to the
  existing reference slopes.
\item Subtract the \cog estimates, $c_x$ and $c_y$, from the existing
  reference slopes.
\end{enumerate}

If a matched filter is already in use, i.e.\ the matched filter is to
be updated, an additional step is required, equal to the reverse of
the above, i.e. adding the \cog of the original image and subtracting
the matched filter estimate of the original image.  We note that a
similar approach has been successfully taken for correlation wavefront
sensing \citep{basden14}.

\subsection{On-bench testing}
We used the CANARY laboratory sky simulator (which uses LED light
sources and two rotating phase screens to mimic the sky) to confirm
that the matched filtering tools were operating as expected, dithering
the sources, and building the matched filters.  The \wfs
noise (due to shot noise and readout noise) was seen to be reduced by
more than a factor of four (from 400~nm to 90~nm), reducing total
wavefront error (in the first 36 Zernike modes) from 950~nm to 750~nm
\rms, for the configuration in which it was used, confirming that the
matched filter was working.  To compute \wfs noise, we
compute noise variance of the wavefront slope measurements, and
propagate this through a theoretical Zernike reconstruction matrix,
and sum the first 36 terms, as is done by \citet{2014A&A...569A..16V}.

\subsection{On-sky measurements}
In this section, the on-sky performance is given in term of Strehl
ratio as directly measured in the H-band infrared images recorded by
the on-axis CANARY science camera.   The \ao loop is
always closed on the on-axis elongated \lgs for the high order
correction while the tip and tilt correction is provided by the CANARY
on-axis Truth \wfs. The \lgs slope estimation
algorithm was alternated between \cog and matched filtering.

\begin{figure}
  \includegraphics[width=\linewidth]{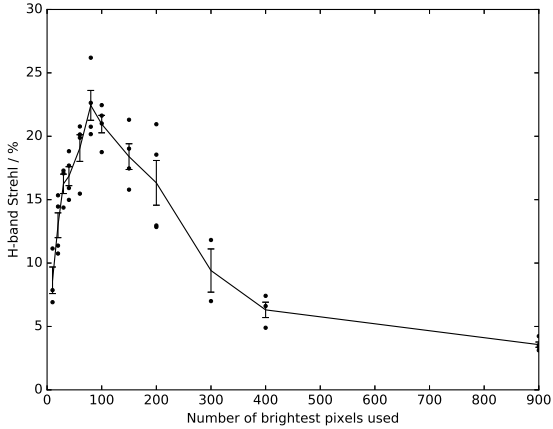}
  \caption{A figure showing AO performance as a function of number of
    brightest pixels chosen for the LGS WFS when using a \cog algorithm.  Closed-loop H-band Strehl ratio is shown, with the LGS
    used for high order correction, and an on-axis NGS for tip-tilt
    correction.  The line with error bars is the standard deviation computed from the
    available information (shown) for each number of pixels.  The LGS
    had about 13~arcseconds of elongation.}
  \label{fig:brightest}
\end{figure}

The \cog algorithm used as the reference for these performance
comparisons was well optimised: we use the brightest pixel selection
algorithm \citep{basden10} to significantly reduce the effect of
detector readout noise, with about 90 brightest pixels being selected,
this number having been chosen for best performance, as shown in
Fig.~\ref{fig:brightest}.  In this case, the 90 brightest pixels are
used for calculation of the \cog, after a threshold equal to the value
of the 91\textsuperscript{st} pixel is applied.  Reference slopes for
both the \cog and matched filter were measured on-sky, averaging 10000
frames of data (66~s) before using each algorithm.

\subsubsection{16th September 2016}
Initial on-sky measurements were taken on the night beginning on 16th
September 2016.  The \lgs was elongated to about 13~arcseconds as
measured by the \wfs, and the elongation was aligned with the
horizontal axis of the \wfs.  The amplitude of the LGS dither was set to 0.1~arcseconds diameter, \ignore{About 0.32 arcsec per volt - and
  dither amp was 500} and 1000 frames of data were used (250 at each
dither location) during recording of the image gradient, $G$, and a
separate 1000 frames used to measure $I_0$ (without dithering).  The
frame rate was 150~Hz, so 1.6~s of data was recorded for each dither
location.  \ignore{The \ao loop was engaged on the \lgs \wfs, with tip-tilt
correction provided by an on-axis \ngs (using \cog slope
estimation). The \lgs slope estimation algorithm was alternated between \cog and matched filtering.  Reference slopes for both the \cog and matched filter were
measured on-sky, averaging 10000 frames of data (66~s) before using
each algorithm.}

These results were measured close to dawn, when sky background was
increasing.  Eventually, the \ngs tip-tilt signal was insufficient to
provide good correction, so we did not record further data.  We
interleaved \cog and matched filter measurements.

\begin{figure*}
  \includegraphics[width=\linewidth]{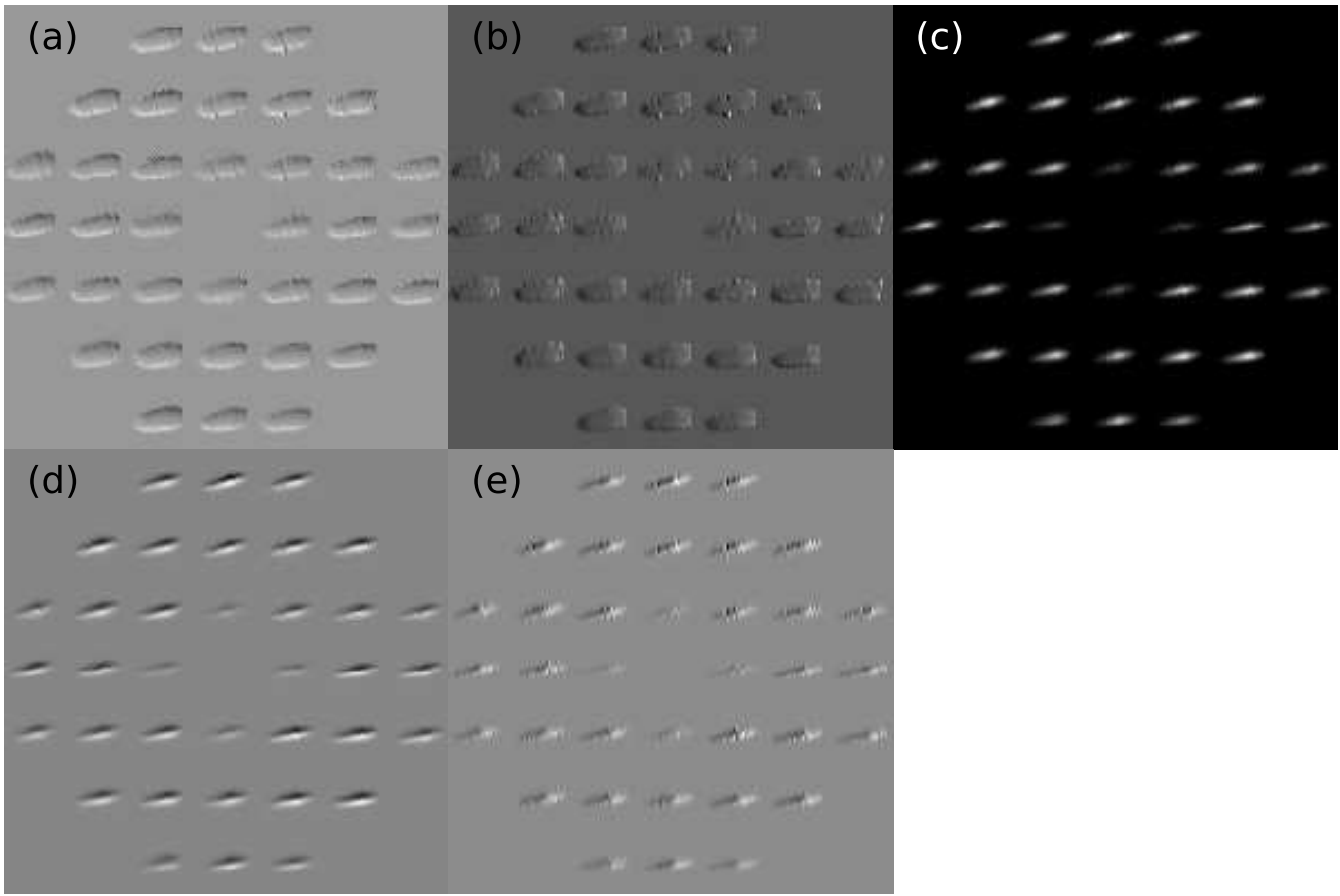}
  \caption{A figure showing, for the $7\times7$ sub-apertures of the
    CANARY phase D \lgs, (a) an on-sky matched filter x component,
    (b) the matched filter y component, (c) $I_0$,
    (d) $G_x$,  and (e) $G_y$.  We note that we had some trouble with
    detector voltage levels, resulting in a few streaks in the image.
  This is manifested within the matched filter.}
  \label{fig:lgsmatchedfilter}
\end{figure*}

A single matched filter was used that for the duration of the
experiment on this night.  The matched filter computed from the data is shown in
Fig.~\ref{fig:lgsmatchedfilter}, along with the components used to
make it, $I_0$, $G_x$, and $G_y$.  We note that a few streaks are seen
in these images due to problems with detector voltage levels
(currently being repaired).

\begin{figure}
  \includegraphics[width=\linewidth]{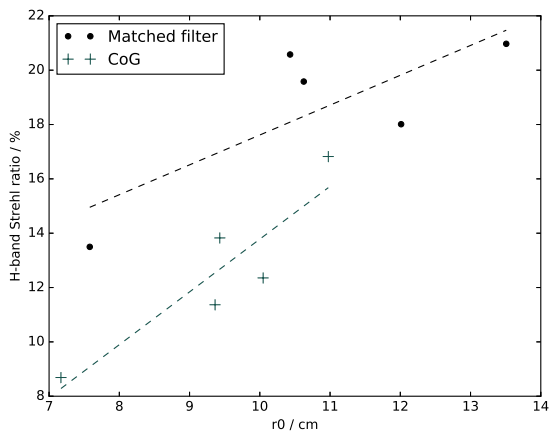}
  \caption{A figure showing H-band Strehl ratio for CoG and matched
    filtering measured on the night of 16th September.  Data are
    plotted as a function of Fried's parameter, $r_0$, which is
    related to seeing.   A 2~s exposure time is used for the science
    images, for a star with a V-band apparent magnitude of 9.47.  The \lgs
    \wfs is recording approximately 3000 photons per sub-aperture per
    frame.  The matched filter and corresponding reference slopes were
    recorded preceeding acquisition of the data.}
  \label{fig:res16th}
\end{figure}

Fig.~\ref{fig:res16th} shows measured \ao system performance with and
without matched filtering.  It should be noted that due to atmospheric
variability, we interleave measurements, so that trends in seeing with
time can be somewhat mitigated.  We present these results as a
function of $r_0$, which is computed from the average value computed
using the pseudo-open-loop slope measurements from the three off-axis
\ngs wavefront sensors using standard CANARY tools
\citep{2014A&A...569A..16V}.  It is evident that the matched filter is
providing significantly better performance than the \cog algorithm.

\ignore{
By using Maréchal's approximation \citep{marechal} linking wavefront
error to Strehl ratio, we can approximate the reduction in wavefront
error associated with use of the matched filter.  We note that these
estimates are only approximate, since Maréchal's approximation is
usually only valid for higher Strehl ratios.  However, in this case,
it gives some idea of improvement, which we can then compare with
other studies.  For 11~cm seeing, Strehl ratios of about 15.7\% and
18.7\% are recorded for \cog and matched filter algorithms
respectively.  }

\subsubsection{17th September 2016}
\label{sect:cogOptimzed}
Further investigation of matched filtering performance was carried
out, with the \lgs spots elongated to about 13~arcseconds, at an angle
of about 10~degrees from the horizontal axis of the \wfs.  This time,
the matched filter was built using a reference image ($I_0$) that was
computed while dithering, using a shift-and-add of the Shack-Hartmann
images, based on the \cog of the spots.

\begin{figure}
  \includegraphics[width=\linewidth]{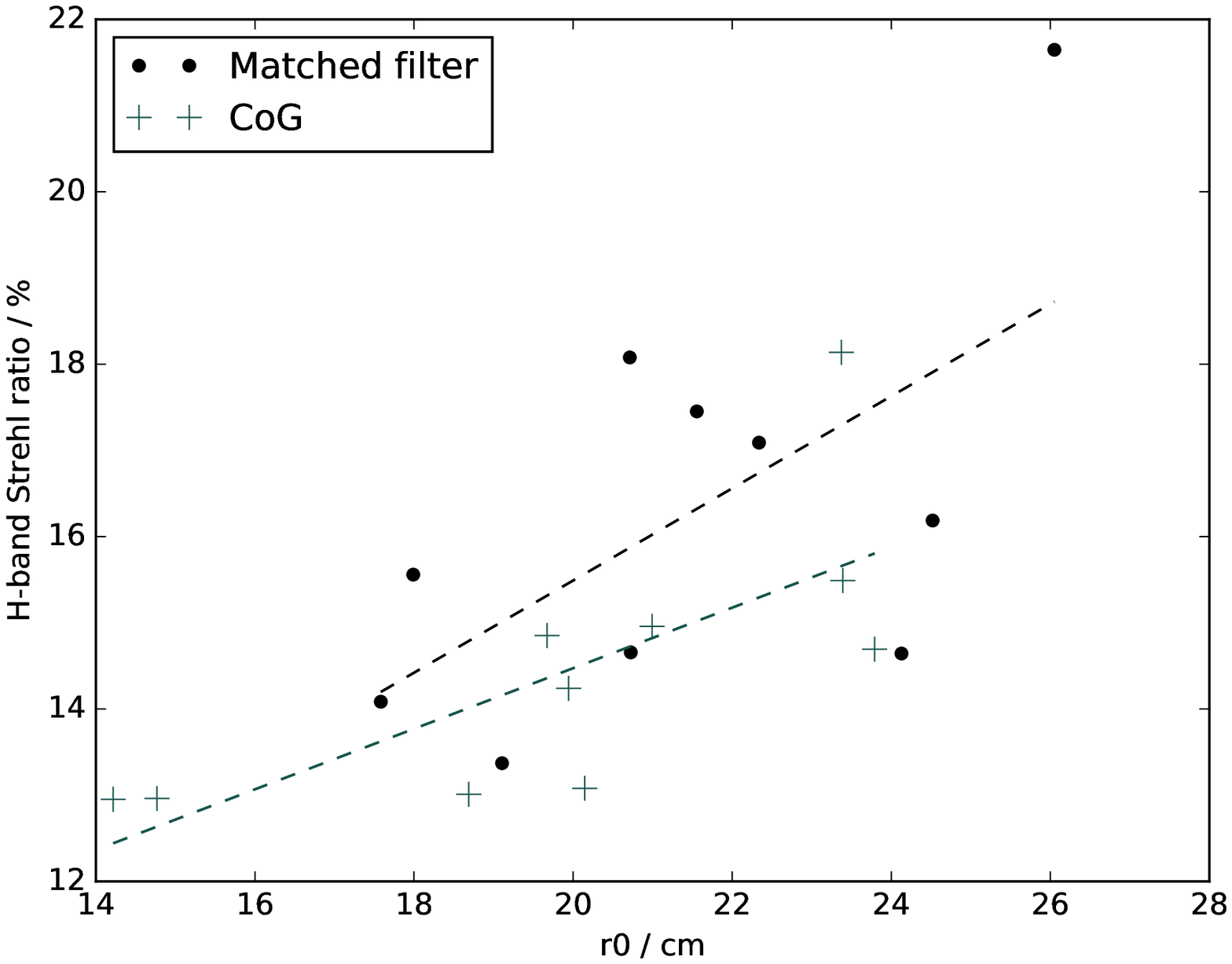}
  \caption{A figure showing H-band Strehl ratio as a function of
    Fried's parameter for the different
    centroiding modes measured on the night of 17th September.  Black circles represent matched filtering,
    while crosses are for centre of gravity.  Mean Strehl values are given in
    the legend.  The same target was used as on the previous night.}
  \label{fig:plot17th}
\end{figure}

Strehl ratio was recorded using the CANARY infrared camera, at H-band,
and Fig.~\ref{fig:plot17th} shows these results.  There is significant
scatter due to variable seeing conditions.  However, matched
filtering can be seen to deliver performance improvements compared to
\cog, increasing the mean Strehl ratio by a factor of greater than 1.1.

\subsubsection{19th September 2016}
When building a matched filter, the dither amplitude used will affect
performance.  We therefore investigated three different dither
amplitudes, setting it first to 0.2~arcseconds diameter, then to
0.1~arcseconds (as on previous nights) and finally to 0.05~arcseconds.
Again, the \lgs spots were elongated to about 13~arcseconds at an
angle of about 10~degrees.  In this case, the reference image, $I_0$
was obtained from averaging 1000 image frames without dithering (and
without shift-and-add).  As before, reference slopes were measured
on-sky before the \ao loop was closed for each matched filter and \cog
acquisition, in this case, with each dither amplitude.

\begin{figure*}
  \includegraphics[width=\linewidth]{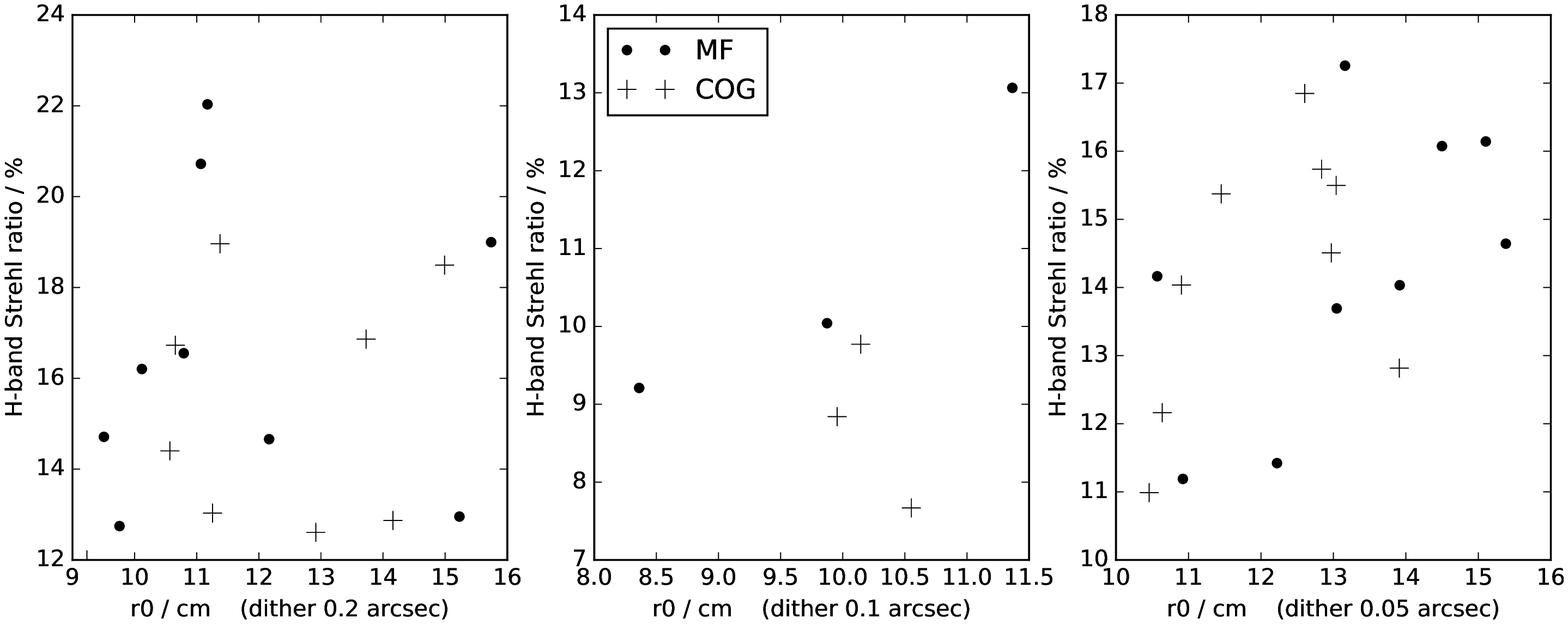}
  \caption{A figure showing AO performance (H-band Strehl) for
    different closed-loop data sets measured on the night of 19th
    September using matched filtering and \cog algorithms, as a
    function of Fried's parameter. The mean of
    these measurements are given in table~\ref{tab:meanres}.
    In this case, the science target has an
    apparent magnitude of 8.98 and a 3~s exposure is used.  The
    matched filter, and reference slopes for both \cog and matched
    filter were recorded immediately before the start of each dataset.
    The dither amplitude used to create the matched filter is, from
    left to right, 0.2~arcseconds, 0.1~arcseconds and 0.05~arcseconds.}
  \label{fig:plot19th}
\end{figure*}

Fig.~\ref{fig:plot19th} shows the \ao performance obtained.  Again it
can be seen that the matched filter algorithm gives consistently
better performance than the \cog, except for the case with a dither
amplitude of 0.05~arcseconds, where the performance gain is marginal.  The \cog and matched filter
measurements are given as a function of Fried's parameter, $r_0$ so as
to account for variable atmospheric seeing.
We note that there is no algorithmic difference between the different
CoG points in each figure,
they were just recorded at different times (interleaved with the
corresponding matched filter measurements), and hence different
atmospheric conditions, using different sets of
on-sky reference slopes.  The mean matched filter measurements
(table~\ref{tab:meanres}) show an
improvement over the corresponding \cog measurements, though we note
that due to differing seeing, computing a mean Strehl ratio does not
give an accurate performance estimate; rather it serves to give us a
rough idea of relative performance.  From these measurements, and from
Fig.~\ref{fig:plot19th}(c), it is evident that a dither of
0.05~arcseconds is too small for us to build a matched filter (this
equates to 7\% of a pixel).

\begin{table}
  \begin{tabular}{llll}
    Dither & Matched filter & \cog & Improvement \\
    amplitude & Strehl / \% & Strehl / \% \\\hline
    0.2~arcsec &16.6$\pm$ 1.1& 15.1$\pm$0.9 & 10\%\\
    0.1~arcsec &10.8$\pm$1.2& 8.8$\pm$0.6 & 23\%\\
    0.05~arcsec &14.3$\pm$ 0.7& 14.2$\pm$ 0.6 & 1\%\\ \hline
  \end{tabular}
  \caption{A table showing matched filter and \cog performance for
    different dither amplitudes.  The relative performance improvement when
    using matched filtering is also given.}
  \label{tab:meanres}
\end{table}

\subsubsection{On-sky results summary}
We have explored matched filtering performance with \eelt scale extended \lgs
Shack-Hartmann images over several nights of operation, and in each
case, we have found that matched filtering leads to \ao performance
improvements when compared with \cog.

We note that in our case, because all sub-apertures were well
elongated, the gain in \wfs performance using the matched
filter was evident.  For an \elt, with elongation varying more
significantly across the \wfs, the total performance gain
may be smaller, though still significant. We also note that the LGS elongation was varying during the night and from night to night, following the sodium layer thickness variations \citep{2010A&A...511A..31M}. 

\subsection{Open-loop considerations}
Our tests of matched filtering have been implemented using a
closed-loop \ao system, and so measured performance may be less
affected by non-linearity effects than would be seen using an
open-loop \ao system, since the \ao loop helps to maintain the
Shack-Hartmann spots close to their reference positions.  Therefore,
performance with an open-loop \ao system may not be so positive since
spot motion will be larger, and spots may travel outside the linear
range of the matched filter.  In this case, either a two-step
algorithm is required (using an initial \cog estimate to find the
approximate location to apply the matched filter to), or the use
adaptive sub-aperture windowing allowing the sub-aperture to track the
spot motion.

There are no proposed \elt instruments that are entirely open-loop:
although the proposed MOSAIC instrument \citep{Hammer2014} on the
\eelt has an open-loop \moao system, it operates with a closed-loop
\glao correction, and therefore, a significant amount of spot motion
will be mitigated \citep[for likely atmospheric turbulence
  profiles,][]{35layer,2010MNRAS.406.1405O}.  Therefore matched
filtering may still be appropriate here.

\subsection{Future work}
Although we have demonstrated \lgs \ao operation of a matched filter algorithm
on-sky, there remains further much work to do, which we will attempt during
future CANARY on-sky observing runs.  \ignore{Demonstration of matched
filtering for partial open-loop operation, using one closed-loop and
one open-loop \dm will verify this technique for use with the MOSAIC
\eelt instrument (which is the only proposed \elt instrument with
open-loop \dms).}  Building the matched filter during \ao loop
operation, and determination of pixel scale change, is essential
to improve \ao observational efficiency.

Additionally, successful implementation of on-line reference slope
calculation is also necessary: for the measurements presented here, we
compute reference slopes for the matched filter algorithm on-sky, by
averaging many frames of slope measurements (typically 10000) to
average atmospheric turbulence.  These slopes are then added to the
reference slopes required to correct non-common path aberrations.
This method does not represent efficient use of telescope time unless
slope measurements are monitored and updated continuously while in
operation.  Therefore a method to determine new matched filter
reference slopes using the matched filter and initial reference slopes
(either from a \cog or previous matched filter) is required.

A further in-depth study of matched filter performance for different
\lgs sodium layer profiles and elongations is also necessary, with
the improvements in wavefront error being calculated when the matched
filter algorithm is used. Moreover we will compare the matched filtering algorithm with other slope estimation methods in greater depth.
The comparison with other  slope estimation methods requires a sufficient number of on-sky observations to be 
statistically meaningful, hence more observing time.

\section{Conclusions} 
We have successfully implemented and demonstrated a noise-optimal
matched filtering algorithm for determination of local wavefront
gradients using a Shack-Hartmann sensor.  On-sky testing was performed
using the CANARY \ao demonstrator instrument on the William Herschel
Telescope together with the ESO Wendelstein laser guide star unit. 

We find that matched filtering can deliver \ao performance
improvements compared with the conventional centre of gravity
algorithm, and we measure improvements in H-band Strehl ratio of up to
about 10\% (i.e. the Strehl improves by a factor of 1.1).  

We therefore conclude that a matched filtering algorithm is well
suited for \ao systems with elongated Shack-Hartmann spot patterns,
providing better performance than the commonly used \cog algorithm.
Our demonstration represents a significant risk mitigation though
further work is required. Direct comparison with other methods, such as correlation, also have to be tested.  

\section*{Acknowledgements}
This work is funded by the UK Science and Technology Facilities
Council, grant ST/K003569/1, and a consolidated grant ST/L00075X/1.
The ESO participation and installed infrastructure is funded under the ESO Technology Development Program.
This work was partially supported by the European Commission (Fp7
Infrastructures 2012-1, OPTICON Grant 312430, WP1).  This work is also
supported by the European Southern Observatory.  CANARY is a visitor
instrument to the William Herschel Telescope, operated by the Isaac
Newton Group in the Spanish Observatorio del Roque de los Muchachos of
the Instituto de Astrofisica de Canarias.

\bibliographystyle{mn2e}

\bibliography{mybib}
\bsp

\end{document}